\documentclass[preprint2]{aastex}

\slugcomment{Not to appear in Nonlearned J., 45.}

\shorttitle{NGC 6231}
\shortauthors{Sung et al.}

\begin{document}

\title{The Initial Mass Function and the Surface Density Profile of NGC 6231}

\author{Hwankyung Sung}
\affil{Department of Astronomy and Space Science, Sejong University,
    98, Kunja-dong, Kwangjin-gu, Seoul 143-747, Korea\footnote{Visiting Fellow,
    Research School of Astronomy and Astrophysics, Australian National
    University}}
\email{sungh@sejong.ac.kr}

\author{Hugues Sana}
\affil{Astronomical Institute `Anton Pannekeok', Amsterdam University, Science
Park 904, 1098 XH, Amsterdam, The Netherlands}
\email{H.Sana@uva.nl}

\and

\author{Michael S. Bessell}
\affil{Research School of Astronomy and Astrophysics, Australian
National University, MSO, Cotter Road, Weston, ACT 2611, Australia}
\email{bessell@mso.anu.edu.au}

\begin{abstract}
We have performed new wide-field photometry of the young open cluster NGC 6231
to study the shape of the initial mass function (IMF) and mass segregation. 
We also investigated the reddening law toward NGC 6231 from optical to 
mid-infrared color excess ratios, and found that the total-to-selective 
extinction ratio is $R_V$ = 3.2, which is very close to the normal value. 
But many early-type stars in the cluster center show large
color excess ratios. We derived the surface density profiles 
of four member groups, and found that they reach the surface density 
of field stars at about 10$'$, regardless of stellar mass. 

The IMF of NGC 6231 is derived for the mass range  0.8 -- 45 M$_\odot$.
The slope of the IMF of NGC 6231 ($\Gamma = -1.1 \pm
0.1$) is slightly shallower than the canonical value, but the difference 
is marginal. In addition, the mass function varies systematically, and is
a strong function of radius - it is is very shallow at the center, 
and very steep at the outer ring suggesting the cluster is mass segregated. 
We confirm the mass segregation for the massive stars ($m \gtrsim$ 8 M$_\odot$)
by a minimum spanning tree analysis. Using a Monte Carlo method, we estimate
the total mass of NGC 6231 to be about $2.6 (\pm 0.6) \times 10^3$ M$_\odot$.
We constrain the age of NGC 6231 by comparison with evolutionary isochrones.
The age of the low-mass stars ranges from 1 to 7 Myr with a slight peak
at 3 Myr. However the age of the high mass stars depends on the adopted
models and is 3.5 $\pm$ 0.5 Myr from the non- or moderately-rotating models of
Brott et al. as well as the non-rotating models of Ekstr\"om et al.
But the age is 4.0 -- 7.0 Myr if the rotating models of Ekstr\"om
et al. are adopted. This latter age is in excellent agreement with the time scale of ejection 
of the high mass runaway star HD 153919 from NGC 6231, albeit the younger age cannot 
be entirely excluded.

\end{abstract}

\keywords{stars: formation -- stars: pre-main sequence --
open clusters and associations: individual (NGC 6231)}

\section{INTRODUCTION}

The young open cluster NGC 6231 is considered as the core cluster of the Sco OB1
association \citep{phyb90} and is one of the brightest open clusters in
the southern hemisphere. Although there is a small amount of differential
reddening \citep{sbl98,rcb97}, the reddening of the inner 10$'$ region is
relatively constant and low. NGC 6231 is a rare case of an exposed 
young open cluster.
In addition, there are about 150 O, B, and A main sequence (MS) or post-MS
stars within 10$'$ from the center of NGC 6231, and therefore NGC 6231
is a good target for the study of dynamical evolution of open clusters
\citep{rm98}. In most cases there are only a handful of massive O-type
stars in nearby open clusters (the Orion Nebula Cluster or NGC 2264) and as 
a result, the initial mass function (IMF) of the massive part of these clusters
inevitably suffers from low-number statistics.
NGC 6231 is richer in massive stars than other nearby young open clusters, and is
relatively closer and more transparent than young compact clusters such as NGC 3603
\citep{sb04}. In this respect, NGC 6231 is a good target for the study
of the stellar IMF, especially in the massive part.

One of hot issues in cluster research is the origin of mass segregation
in open clusters. The failure of the classical approach was well
tested in \citet{mb09}, and they concluded that several young open clusters,
including the Orion Nebula Cluster and NGC 6231, are too young to be dynamically
relaxed. They also concluded that strong primordial mass segregation is 
unlikely to be dynamically erased over the first few Myr of cluster evolution.
The breakthrough in the so-called ``primordial mass segregation'' in young
open clusters was made by \citet{mvpz07}. They showed that mergers
of small clumps, that are initially mass segregated, lead to larger systems
that inherit the progenitor clumps' segregation. \citet{rja09b,rja10} performed
numerical simulations of cool, clumpy clusters, and showed that subvirial,
fractal clusters are often dynamically mass segregated on timescales far 
shorter than that expected from simple dynamical models. \citet{rja10} showed that the 
formation of Trapezium-like massive multiples is highly probable in such 
a cool, clumpy cluster at a very early phase. They also showed the possibility 
of ejection of massive stars due to the dynamical interaction between massive
stars during the core collapse phase of the cluster. In addition, 
\citet{rja09a} devised an objective method to calculate the degree of 
mass segregation based on the minimum spanning tree method. One of the important
aims of this work is to study the extent of mass segregation in NGC 6231.

\citet{sbl98} (hereafter SBL98) performed $UBVRI$ \& H$\alpha$ photometry
and found several
pre-main sequence (PMS) stars and PMS candidates. But the number of PMS
stars with H$\alpha$ emission was far smaller than the number of low-mass
PMS stars expected from the number of massive MS and
post-MS stars in the cluster. They suggested possible explanations
for the deficit of PMS stars with H$\alpha$ emission, such as past supernova 
explosions or strong stellar winds from massive stars in the cluster.
Later \citet{bvf99} undertook a statistical approach and confirmed the
existence of a large number of PMS stars in the cluster. From analysis of the 
180-ks {\it XMM-Newton} campaign of NGC 6231, \citet{sgrsv} found 610 X-ray
sources in the field of view and confirmed that many of these X-ray sources
are bona-fide low-mass PMS stars in NGC 6231 \citep{sana07}. Recently 
\citet{damiani09} detected more than 1600 X-ray point sources from
the 120-ks {\it Chandra} ACIS observation, and confirmed the \citet{sana07}
results. In addition, they analyzed the X-ray spectra of about 500 X-ray
sources without optical or near-IR counterparts, and found unusually hard
spectra for these sources. These objects could be low-mass PMS stars ($V-I 
\gtrsim 2.0$) showing high coronal temperatures (see Fig. 1 of \citet{sbs08}).

Another interesting issue for NGC 6231 concerns past supernovae explosions in the 
cluster. The massive star content, the IMF, and the age of
NGC 6231 strongly suggest the likelihood of at least one supernova. \citet{msb74} 
studied the O6.5Iaf star HD 153919, the optical companion of the X-ray binary, 
2U1700-37, and suggested from the reddening and brightness of the star, that
it may be at the same distance as Sco OB1. Later \citet{ankay01} suspected 
that Sco OB1 may be the origin of this runaway high-mass X-ray binary using the high-quality
proper motion data from {\it Hipparcos}. From the optical polarization 
observation of 35 stars in the cluster, \citet{fmmvbv} found a curious 
semicircular pattern of polarization angles of some stars in the center of 
NGC 6231. 
They interpreted it as a fingerprint of a past supernova explosion. But they
seem not to have realized the existence of the runaway high-mass X-ray binary 
HD 153919. Alternatively, \citet{black08} proposed the semicircular pattern of
polarization angles as the remnants of the historical nova in 1437 recorded
in the old Korean Veritable Records of King Sejong {\it Sejong Sillok}.

\citet{dr96} studied the binary frequency of B type stars in NGC 6231 and derived
a minimum fraction of 52\%. Recently \citet{sana08} studied the binary fraction
of O type stars in the cluster and obtained a minimum fraction of 0.63.
The fraction of binaries, as well as the mass ratio in binary systems, could
be very important subjects of study (see \citet{sana12}), because we still have insufficient
information on the mass function of secondaries in binary systems
(see \citet{wkm09} for its influence on the mass function). In this respect
NGC 6231 could be a good target because it is relatively closer and therefore
it is easy to obtain high quality spectra with medium-size telescopes.

New wide-field $UBVI$ and H$\alpha$ photometry of NGC 6231 is presented 
and compared in section 2. Based on the new photometric data, several
photometric diagrams are presented in section 3. The reddening law toward 
NGC 6231, reddening, distance, and surface density profiles are also derived in
section 3. The Hertzsprung-Russell diagram (HRD) of NGC 6231 is constructed
in section 4. The IMF and age of NGC 6231 is derived from the HRD by comparing 
the observed data and stellar evolution models in this section. The mass
segregation in NGC 6231 is investigated in the same section. Discussion on 
the runaway O6.5Iaf star HD 153919 and the origin of mass segregation is 
presented in section 5. The summary is in section 6.

\section{OBSERVATIONS AND DATA REDUCTION}

$UBVI$ and H$\alpha$ CCD photometry of NGC 6231 was performed on 2000 June 22 
(NW \& SW regions) and 25 (NE \& SE regions)
at Siding Spring Observatory with the 40 inch (1 m) telescope (f/8) and 
a thinned SITe 2048 $\times$ 2048 CCD (24 $\mu m$ pixels). 
The filters used were the same as those used for the observation of NGC 2244
\citep{ps02}. Two sets of exposure times were used in the observations --
long: 60 s in $I$, 150 s in $V$, 300 s in $B$, 600 s in $U$ and H$\alpha$ --
and short: 2 s in $I$ and $V$, 3 s in $B$, 30 s in $U$ and H$\alpha$.
The seeing was good on 2000 June 25 (about $1.''4$ in $V$ long-exposed 
images), but was bad on 2000 June 22 ($2.''2$ -- $3.''2$). The seeing strongly 
affects the photometry of faint stars - completeness and limiting magnitude. 
The limiting magnitude in $V$ on 2000, June 22 ($V_{limit} \approx 20.5$ mag) 
is more than 0.5 mag brighter than that on 2000, June 25 ($V_{limit} \gtrsim 
21.0$ mag).

All the preprocessing, such as the overscan correction, bias subtraction, and
flat fielding, was done using the IRAF/CCDRED package. Instrumental
magnitudes were obtained using IRAF/DAOPHOT via point spread function fitting
for the target images and via simple aperture photometry for standard stars.
All the instrumental magnitudes were transformed to the standard $UBVI$
system using SAAO E-region standard stars, E5 (E5-R, -O, -T), E6 (E6-U, 61, -Y,
-W), and E7 (E7-W, 37, -c, -X) (see \citet{sb00} for details). We observed
E5 (and E7 on 2000, June, 25) region(s) twice at high (airmass $\approx$ 1.1)
and low altitude (airmass $\approx$ 2.0) to 
determine the atmospheric extinction coefficients and the time variation
coefficients accurately. The primary extinction coefficients, time variation
coefficients of the photometric zero points (unit: magnitude per hour - see 
equations (1) -- (4) in \citet{sbcki08}), and photometric zero points at
midnight were derived from standard
star photometry, and listed in Table \ref{tab_coef}. On 2000 June 22,
the time variation was a function of wavelength, largest in $U$ but zero in $I$.
On 2000 June 25, the time variation in $UBV$ had similar values, but in $I$
it had the opposite sign. The secondary extinction coefficients for $U$ and $B$
were adopted from the mean values used in \citet{sb00}. Although there were
no standard magnitudes in H$\alpha$, we transformed H$\alpha$ to $R$ 
(i.e. pseudo-$R$) so as to calculate the H$\alpha$ index \citep{scb00,ps02}.

\placetable{tab_coef}

\begin{figure}
\epsscale{1.0}
\plotone{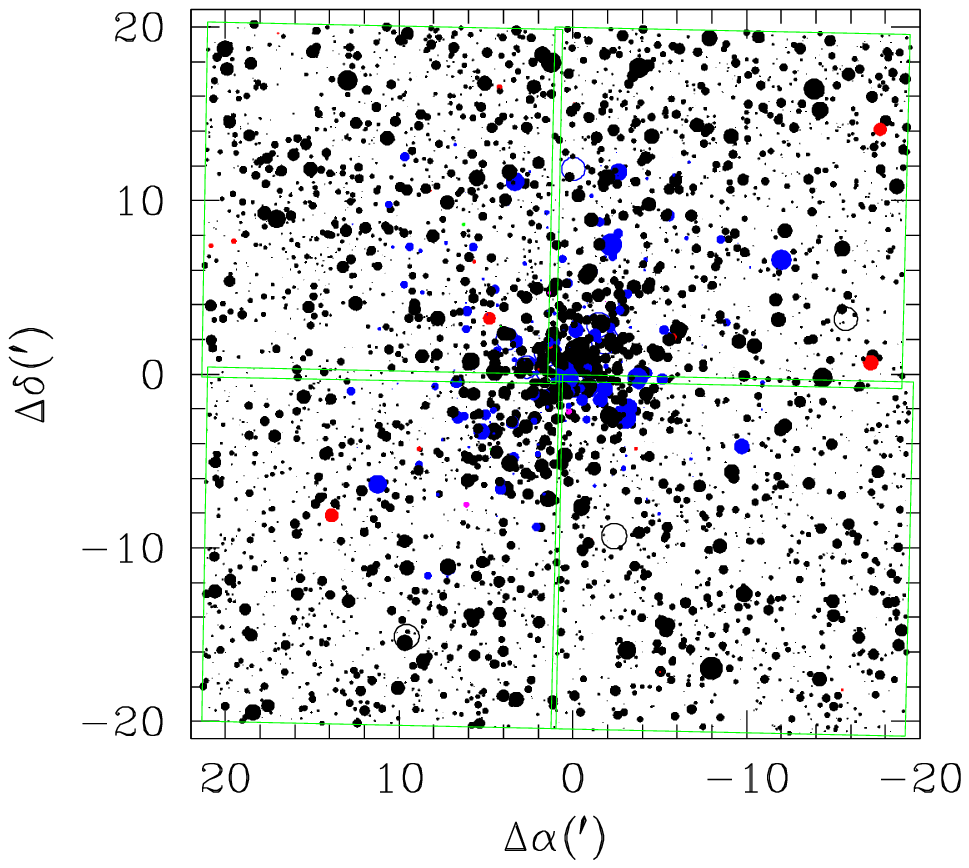}
\caption{Finder chart of NGC 6231 for the stars brighter than $V$ = 17. 
The size of the dots is proportional to the brightness of the star. Large open
circles represent the position of bright stars (data from SIMBAD database). 
The position of stars is relative to the adopted center of NGC 6231 ($\alpha$(
J2000) = 16$^h$ 54$^m$ 11.$^s$70, $\delta$(J2000) = -41$^\circ$ 50$'$ 20.$''$3).
Large squares represent four fields of view of the SSO SITe 2048 $\times$ 2048 CCD.
\label{figmap} }
\end{figure}

A total of 27,325 stars were measured in about a $40' \times 40'$ area centered
on NGC 6231, and are presented in Table \ref{tab_data}. For 12 stars in Table 
\ref{tab_data} that were saturated in our images we have substituted data from
SIMBAD\footnote{\url{http://simbad.u-strabg.fr/simbad}}, 
WEBDA\footnote{\url{http://www.univie.ac.at/webda/}} (Open cluster database), or SBL98.
Data for some stars missed in the new photometry were taken from SBL98. 
These data are marked with an ``*'' after the identification number in the first
column. We also listed 2MASS identification of the objects in the 2MASS
point source catalogue \citep{2mass} in the 16th column. The optical counterparts
of X-ray sources \citep{sgrsv} are identified, and marked as ``X'' (the closest 
or the only objects within the search radius of 3.$''$5) or ``x'' (star within
3.$''$0, but second closest star). We also identified  H$\alpha$ emission stars
(see section 3.1), and these are marked as ``H'' (strong H$\alpha$ emission stars)
or ``h'' (weak H$\alpha$ emission candidates). The membership information
is in the 17th column. A total of 522 X-ray emission stars (``X'' in column 17)
and 70 X-ray emission candidates (``x'') are identified among 610 X-ray sources
\citep{sgrsv}. Among them, 11 stars are X-ray emission stars with H$\alpha$ 
emission (``+''), and 7 stars are X-ray emission stars with weak H$\alpha$ 
emission (``-''). Other information such as duplicity, variability, ID in
SBL98, spectral type, and other identifications are listed in subsequent
columns. The finder chart based on the new photometry is shown in Figure \ref{figmap}.

\placetable{tab_data}
\placefigure{figmap}

\begin{figure}
\epsscale{0.5}
\plotone{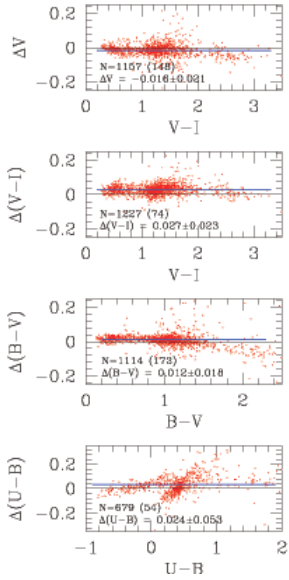}
\caption{Comparison of the photometry. The difference $\Delta$ is in the sense
of our new photometry minus \citet{sbl98} for $V \leq$ 17 mag ($V$, $V-I$, 
and $B-V$) and $V \leq$ 16 mag ($U-B$). Our new photometry is slightly 
brighter in $V$, slightly redder in $V-I$ and $B-V$, but systematically 
bluer in $B-V$ for red stars. The difference in $U-B$ is somewhat larger 
and non-linear. See text for details. 
\label{figsbl} }
\end{figure}

The $UBVI$ magnitude and colors from this photometry were compared with
the previous photoelectric and CCD photometry in Table \ref{tab_comp}.
Although the agreement in zero points was quite good, the scatter was somewhat
large because most photoelectric photometric studies were limited to bright
stars and many of these bright stars are variables or optical doubles.
We compared our new data with SBL98 along with the relevant colors in
Figure \ref{figsbl}. The major difference between the two sets of data was in the standard
stars used. Our new photometry, relative to SBL98, was slightly brighter in $V$
and slightly bluer in $B-V$ for redder stars. Our new $V-I$ was slightly redder,
but with no evident color dependency. For $U-B$, although the average difference
is close to zero, there was a strong color-dependency. Such a difference is
caused by the non-linear correction term  in the transformation to the
Landolt's $U$ system (see \citet{sb00} for details) which is related to
the size of the Balmer jump. 
In this respect (see the bottom panel of Figure \ref{figsbl}) we can understand
the difference between our new $U-B$ and those from photoelectric photometry,
i.e. most stars observed with photoelectric photometry are brighter and bluer,
and therefore our new $U-B$ is slightly bluer for blue stars.

\placefigure{figsbl}
\placetable{tab_comp}

\section{PHOTOMETRIC DIAGRAMS AND SURFACE DENSITY PROFILES} 

\subsection{Color-Color Diagrams and Reddening}

\begin{figure}
\epsscale{0.5}
\plotone{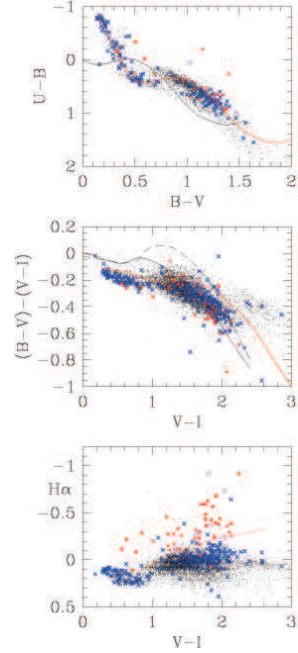}
\caption{The color-color diagrams of NGC 6231. (Top panel) the $U-B$ versus 
$B-V$ diagram for $V \leq$ 16.5 mag. (Middle panel) the $B-V$ versus $V-I$ 
diagram for $V \leq$ 18 mag. (Bottom panel) the H$\alpha$ index versus $V-I$ 
diagram for $V \leq$ 18 mag. Thin and thick solid lines in the upper two 
panels represent 
the intrinsic and reddened ($E(B-V)$ = 0.47 mag) MS relations, respectively.
The dashed line in the middle panel presents the intrinsic relation of giant stars, 
while the thin solid line in the lower panel represents the H$\alpha$ index 
versus $V-I$ relation for normal MS stars \citep{ps02}, but shifted by 0.2 mag 
in $V-I$. Crosses, asterisks, squares, and triangles denote, respectively, 
X-ray emission stars, X-ray emission stars with H$\alpha$ emission, H$\alpha$ 
emission stars, and H$\alpha$ emission candidates. 
\label{figccd} }
\end{figure}

\placefigure{figccd}

The color-color diagrams of the stars in the observed region are shown
in Figure \ref{figccd}. X-ray emission stars and H$\alpha$ emission stars
are marked by different symbols. The intrinsic and reddened MS relations 
(using the adopted reddening law for the ($B-V, ~ V-I$) diagram) 
are also drawn in the figures.

\begin{figure}
\epsscale{0.5}
\plotone{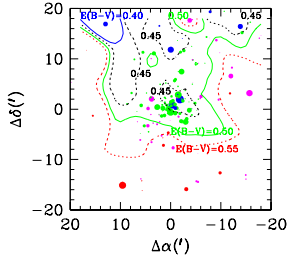}
\caption{The smoothed reddening map of NGC 6231. The lines represent 
the iso-reddening contours smoothed with the scale length of 1$'$. 
The line type and thickness represents different amount of reddening $E(B-V)$
as shown in the figure.
The dot represents the early-type stars used in reddening determination.
\label{figebv} }
\end{figure}

The ($U-B,~ B-V$) diagram is the basic diagram for estimating 
the reddening to early-type stars. The $E(B-V)$ reddening of early-type stars 
in NGC 6231 is between 0.45 and 0.60 mag, but some stars have somewhat 
larger $E(B-V)$. The mean value of $E(B-V)$ is 0.47 mag, which is the same as 
the value obtained by SBL98. The low-mass PMS stars with X-ray and/or
H$\alpha$ emission seem to have nearly the same reddening as the massive early-type stars.
But most of the normal late-type stars are slightly less reddened, and are
presumably foreground field stars in the Sagittarius arm. Unlike the low-mass
PMS stars in NGC 2264 \citep{sbl97}, only a few stars show ultraviolet (UV) 
excesses. This means that the accretion activity of PMS stars in NGC 6231 has
nearly stopped. We determined the reddening $E(B-V)$ of all early-type stars
(We assumed the standard color excess ratio of 0.72 between $B-V$ and $U-B$),
and calculated the spatial variation of reddening as shown in Figure
\ref{figebv}. The dots in the figure represent the early type stars used in 
the reddening determination. As already noted in SBL98, the reddening is very high
in the south, relatively low in the north, and shows a local minimum near 
the center.

\placefigure{figebv}

The ($B-V, ~V-I$) diagram is shown in the middle panel of Figure \ref{figccd}. The reddened MS
relation does not pass through the mean distribution of early-type stars.
This is because the direction of the reddening vector in the diagram is affected 
by the reddening law and the adopted reddening law does not fit the early-type 
stars in the cluster center (see section 3.2 for details).

The (H$\alpha, ~V-I$) diagram is presented for $V \leq 18$ mag in the lower panel 
of Figure \ref{figccd}. The low-mass
PMS stars with H$\alpha$ emission can be selected from the diagram 
\citep{sbl97}. As already noticed in SBL98, there are not many H$\alpha$ 
emission PMS stars in NGC 6231. Most X-ray emission stars do not show excess 
emission in H$\alpha$. We designated stars above the border line \citep{ps02} 
as H$\alpha$ emission stars. In addition, the H$\alpha$ emission stars 
selected in SBL98 or from other studies are also designated as H$\alpha$ 
emission stars. A total of 31 H$\alpha$ emission stars and 34 H$\alpha$ emission
candidates were selected. Among them 18 stars are also X-ray emission stars
with membership ``+'' or ``-''. The star ID 357 (= HD 326324 = Hen 3-1269) 
does not show any appreciable emission in our H$\alpha$ photometry, but 
the star was classified as 
an H$\alpha$ emission star by \citet{henize76}. The stars ID 294 (= Wray 
15-1546, \citep{wray66}) and ID 6017 (= Hen 3-1281) show strong H$\alpha$ 
emission from our H$\alpha$ photometry. ID 2132 (=SBL 125) and ID 4359 (=SBL
667 = Se 166) are two massive H$\alpha$ emission stars near the MS band within
the cluster radius.

\subsection{Reddening Law}

The interstellar reddening law is one of the fundamental parameters used in determining 
the distance to astronomical objects, and is known to be different from 
sight line to sight line in the Galaxy \citep{fitzpatrick09}. They confessed that
there is no universal near-infrared (near-IR) extinction law. We derived similar
relations between $R_V$ and color excess ratio using their equation (5) (the
effective wavelength of {\it Spitzer} IRAC \citep{irac} bands from IRAC 
Instrument Handbook\footnote{\url{http://irsa.ipac.caltech.edu/data/SPITZER/docs/iracinstrumenthandbook/}})
and power $\alpha$ from their Table 4. The results are 

$$ R_V = 1.147 {{E(V-[3.6])} \over {E(B-V) }} - 0.361, ~r=0.998, \eqno{(1)} $$ 
$$ R_V = 1.098 {{E(V-[4.5])} \over {E(B-V) }} - 0.248, ~r=0.999, \eqno{(2)} $$
$$ R_V = 1.065 {{E(V-[5.8])} \over {E(B-V) }} - 0.168, ~r=1.000, \eqno{(3)} $$
$$ R_V = 1.038 {{E(V-[8.0])} \over {E(B-V) }} - 0.101, ~r=1.000. \eqno{(4)} $$

\noindent
We also derived similar relations for the near-IR 2MASS bands, but the data points
showed a large scatter. We decided to use the relations presented in \citet{gv89}
for near-IR $JHK_s$ bands.

SBL98 obtained a slightly higher value of $R_V$ for the stars
in NGC 6231 ($R_V = 3.3$) using the color excess in $V-I$. In this study 
we try to determine the reddening law from the optical $I$ band to the mid-infrared 
(mid-IR) {\it Spitzer} IRAC bands. Firstly one of the authors (M.S.B.) calculated
the synthetic colors of early-type stars in the near-IR 2MASS bands as well as 
in the mid-IR {\it Spitzer} IRAC bands. In addition, we downloaded the {\it Spitzer}
IRAC images from the {\it Spitzer} archive\footnote{\url{http://irsa.ipac.caltech.edu/applications/Spitzer/SHA}}, 
and measured the magnitudes of O and B type stars in the same
way as in \citet{ssb09}. NGC 6231
was observed as a part of the GLIMPSE (Galactic Legacy Infrared Mid-Plane Survey
Extraordinaire) survey, but only part of the cluster
was observed. Our photometric data are compared with those in the GLIMPSE
Point Source Archive (\url{http://irsa.ipac.caltech.edu/data/SPITZER/GLIMPSE/}),
and the results are $\Delta[3.6] = + 0.003 \pm 0.132$ (n = 14), $\Delta[4.5]
= -0.003 \pm 0.071$ (n = 21), $\Delta[5.8] = + 0.038 \pm 0.115$ (n = 14), and
$\Delta[8.0] = +0.032 \pm 0.063$ (n = 21).

\begin{figure}
\epsscale{1.0}
\plotone{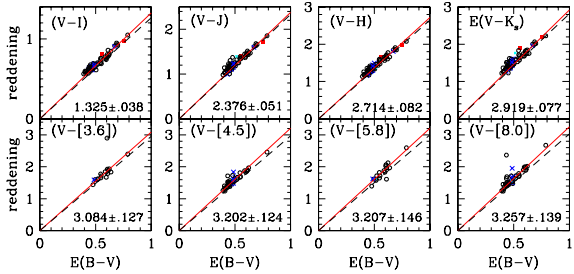}
\caption{The reddening law of NGC 6231 from the optical to mid-IR. The reddening for 
a given color with respect to $E(B-V)$ is shown. The optical and mid-IR data are
from this study, while near-IR data are from 2MASS. The dashed and solid lines represent
the normal reddening law ($R_V = 3.1$) and the reddening law for NGC 6231,
respectively.
\label{figroi} }
\end{figure}

\placefigure{figroi}

The color excess relations between $B-V$ and other colors are presented in
Figure \ref{figroi}. Although many early-type stars with $E(B-V)$ = 0.4 --
0.6 show a large scatter, if we use the stars with $E(B-V)$ $\geq$ 0.6,
the color excess ratios are very close to the normal value. Using the relations
between $R_V$ and color excess ratios, the total to selective extinction ratio 
$R_V$ of the stars in NGC 6231 is 3.22 $\pm$ 0.04 from 8 optical to mid-IR
colors. This value implies that the reddening law toward NGC 6231 is
very close to normal. We adopt $R_V$ = 3.2 for NGC 6231. This value is the same
as the total to selective extinction ratio toward the intermediate-age open
cluster M11 \citep{sblkl99}.

\begin{figure}
\epsscale{0.5}
\plotone{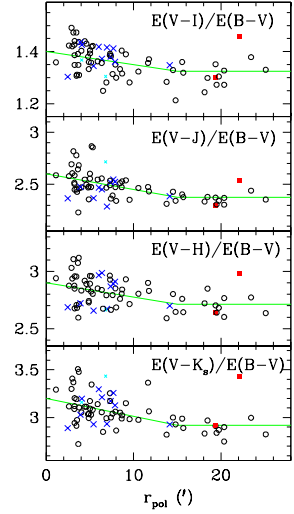}
\caption{Radial variation of color excess ratios. The radial distance in minutes
of arc is measured from the center of polarization angle \citep{fmmvbv}. Squares
and crosses represent stars with H$\alpha$ and X-ray emission, respectively.
The early-type stars in the field region show a nearly normal ratio, but 
those in the cluster show a large scatter. \label{figrrv} }
\end{figure}

\placefigure{figrrv}

In addition, the stars near the cluster center show 
somewhat larger color excess ratios, as seen in Figure \ref{figrrv}.
The abscissa is the distance from the center of the polarization angle \citep{fmmvbv}.
But when the abscissa is distance from the cluster center, the result does
not differ very much. The stars at larger distance from the cluster center
($r_{\rm pol} > 14'$) show very normal ratios, but stars in the inner region 
show large values and a large scatter. The $R_V$ of the inner region from 
4 optical and near-IR bands is 3.43 $\pm$ 0.04. SBL98 proposed that
the deficit of low-mass PMS stars with H$\alpha$ emission and the hole
seen in reddening map are due to the stellar wind from massive stars, WR stars
or past supernova explosion(s). This higher value of the color excess ratios in 
the cluster center may also be related to these effects.

The extra extinction due to the slightly higher value of the color excess ratios
of the stars in the cluster center is about 0.04 -- 0.05 mag in $V$ and 
0.02 mag in $V-I$, if we assume the foreground
reddening toward the Sagittarius arm to be $E(B-V)$$_{fg}$ = 0.2 -- 0.25 mag
\citep{scb00}.
This extra extinction may slightly influence the distribution of stars in 
the middle panel of Figure \ref{figccd} and the upper panel of Figure 
\ref{figcmd}.

\subsection{Color-Magnitude Diagrams and Distance}
  
\begin{figure}
\epsscale{0.5}
\plotone{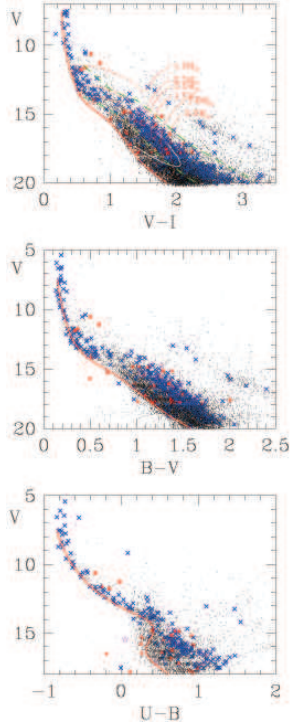}
\caption{The color-magnitude diagrams. (Top panel) the $V-I$ versus $V$ diagram.
The thick solid line represents the reddened ZAMS relation at the NGC 6231 
distance. The thin solid lines are the PMS evolution tracks of \citet{sdf00}, 
while the thin dashed line overlaid on the thick white solid line is the 1 
M$_\odot$ PMS evolution track of \citet{bcah98}. The dashed lines in the 
redder part of the reddened ZAMS line represent the upper and lower limit 
of the adopted locus of PMS stars in NGC 6231. (Middle panel) the $B-V$ versus 
$V$ diagram.  (Bottom panel) the $U-B$ versus $V$ diagram.
The other symbols are the same as Fig. \ref{figccd}.
\label{figcmd} }
\end{figure}

\placefigure{figcmd}

The color-magnitude diagrams (CMDs) of NGC 6231 are presented in Figure 
\ref{figcmd}. From the CMDs at the top left we can easily recognize the well-developed sequence
of early-type MS/post-MS members in the left of each CMD. The zero-age main 
sequence (ZAMS) with mean reddening and adopted distance is over-plotted.
In addition, a few PMS evolution tracks of \citet{sdf00} and the 1 M$_\odot$ 
evolution model of \citet{bcah98} are superimposed.
The empirical color-temperature relations for low-mass stars of
\citet{msb95} and synthetic color - temperature relations of \citet{bcp98}
were employed in the transformation of physical parameters to observational
colors.

Most of the O and early B type stars in NGC 6231 are X-ray emitters. 
In addition, many faint X-ray emission stars are about 3 mag brighter 
than the adopted ZAMS line, and have the typical brightness of
PMS stars in young open clusters. There are not as many X-ray emission
stars at $V-I \approx$ 1.0, but stars in the lower MS and transition zone
are mostly X-ray emission stars. The stars at $V-I \approx$ 1.0 are
field stars at a similar distance to NGC 6231. They are relatively old disk
stars in the Sagittarius arm, and therefore they are mostly inactive in X-rays.
In addition, the small number of X-ray emission stars implies that
the PMS stars evolve quickly at this stage. 

There are five early-type H$\alpha$ emission stars with $V-I <$ 1.0. 
Two of them are within the cluster radius (see section 3.4 for 
the radius of NGC 6231), the brighter three (ID 294 = Wray 15-1546;
ID 357 = HD 326324 = Hen 3-1269; ID 6017 = ALS 3833 = Hen 3-1281)
are outside the radius of NGC 6231. There are two X-ray emission
stars at $V-I \approx$ 0.6 and $V \approx$ 10.5 (ID 1499 = SBL 19 
= HD 326343, B1V; ID 4840 = CD-41 11062B). These two stars are
more reddened ($E(B-V)$ = 0.67 and 0.71, respectively)
than the other early-type stars.
ID 4840 is the fainter component of CD-41 11062, and is an optical double
with ID 4842.

The distance modulus of NGC 6231 can be determined from the distribution of
the distance moduli of individual early-type stars after correcting for interstellar
reddening. The extra extinction mentioned above does not seriously affect 
the estimate of distance to NGC 6231 because the distance modulus of 
individual early-type stars is estimated using the ZAMS
relation in the ($V, ~B-V$) diagram. The distribution shows a peak
at $V_0 - M_V$ = 10.6 -- 11.0 mag, and only a few stars with a larger value.
We adopt the distance to NGC 6231 as $V_0 - M_V$ = 11.0 mag. This value is
in excellent agreement with the value obtained by \citet{sana05}.
Although there is a slight difference in the adopted reddening law, this value
is identical to the value obtained by SBL98 and adopted by
\citet{sgrsv}. The adopted distance modulus and mean reddening were applied
in Figure \ref{figcmd}.

We also checked the completeness of our photometry in the $\Delta V$ ($\equiv
V - V_{\rm bright}$) versus distance from the nearest bright star. 
If the brightness of a star is brighter than or comparable to the local 
brightness around a bright star, the brightness of the star can be measured.
But if the brightness of a star is fainter than or comparable to the local
fluctuation of surface brightness around a bright star, the brightness of 
the star cannot be measured. And therefore the existence or nonexistence
of a faint star in the photometric catalog is a function of the distance 
from a bright star as well as the difference in brightness between two stars. 
The tracing of the avoidance zone 
makes it possible to estimate the area of the incomplete zone due to 
the presence of bright stars. And thus the ratio of the total area and the area
of the incomplete zone could represent the completeness of the photometry.
As the seeing is one of important factor affecting the completeness of 
photometry, the completeness from this method is for the case of bad seeing
of 2000, June, 22.
The completeness of $V$ = 19 mag stars at the cluster center ($r \leq 2'$) is
about 60\%, but is higher than 80\% at $r > 2'$. Although the limiting
magnitude of our photometry is about $V$ = 20.5 mag, the magnitude of over 
80\% completeness, is about $V$ = 19 mag, except for the cluster center 
($r \leq 2'$).
The dashed lines in the $V$ versus $V-I$ diagram are the apparent loci of
PMS stars (X-ray emission stars and H$\alpha$ emission stars) of NGC 6231.
From the completeness test results, our photometry is about 80\% complete 
for PMS stars with mass $\gtrsim$ 0.8 M$_\odot$, except for the cluster
center ($r \leq 2'$).

There is a relatively well-defined sequence of foreground stars in the CMDs. 
They are about 5.5 mag brighter than the adopted ZAMS of NGC 6231.
Several X-ray emission stars can be found in the sequence. The distance
to the foreground objects is about 130 pc. As NGC 6231 is located at the 
south-eastern edge of the Upper Centaurus-Lupus group of the Sco-Cen OB 
association, the stars are likely members of the Upper Centaurus-Lupus moving group.

\subsection{Surface Density Profiles and the Radius of NGC 6231}

\begin{figure}
\epsscale{0.5}
\plotone{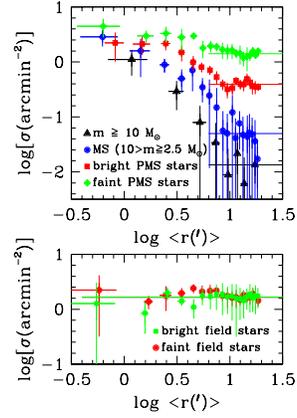}
\caption{Surface density profiles. (Upper) Surface density profiles
of MS stars and stars in the PMS locus of NGC 6231. The horizontal bars
represent the average surface density of field region ($r > 10'$).
(Lower) Surface
density profiles of the field stars above and below the PMS locus of NGC 6231.
The widths of the annuli are 1$'$ for all groups except the most massive group 
($\Delta r = 2'$).
\label{figrho} }
\end{figure}

\placefigure{figrho}

NGC 6231 is the core cluster of the Sco OB1 association, and therefore the radius
of the cluster is not easy to determine because many members of the Sco OB1
association, as well as field stars in the Sagittarius arm, are seen in the same
sight line. In particular, the photometric characteristics of Sco OB1 association
members are very similar to those of the cluster members. 

To determine the radius of NGC 6231, we assigned the membership of each star.
Stars with $V \leq$ 13 mag and $U-B \leq$ 0.0 mag were classified as massive MS
members of NGC 6231. And stars in the PMS locus (the upper panel of Figure
\ref{figcmd}) were classified as PMS members if X-ray emission and/or H$\alpha$
emission is detected, or as PMS candidates if no membership information 
is available. Next we determined the center of NGC 6231 from the geometric 
center of stars with a certain brightness limit. Different brightness limits
gave a slightly different value, but the difference was not large. We 
calculated the geometrical center of the massive MS members with masses greater 
than about 5 M$_\odot$ ($V \approx$ 11 mag) and adopted this as the center
of NGC 6231. The adopted center of NGC 6231 is $\alpha_{\rm J2000} = 16^h ~54^m ~11.^s7, ~\delta_{\rm
J2000} = -41^\circ ~50' ~20''$, which is approximately equidistance from
HD 152248 (O7.5III+O7III(f) - \citet{sana01}), HD 152249 (O9Ib(f) - 
\citet{sana08}), and HD 152270 (WC7+O6V - \citet{hill00}).

We calculated the surface density profiles for 4 member groups - massive MS/post-MS
members ($V \leq$ 10, equivalently $m \gtrsim$ 10 M$_\odot$), intermediate-mass
MS stars ($V$ = 10 -- 13, equivalently $m$ = 10 -- 2.5 M$_\odot$), bright PMS
stars ($V \leq 15$ in PMS locus), and faint PMS stars ($V$ = 15 -- 18 in
PMS locus), and these are shown in Figure \ref{figrho}. In the calculation of
surface density, we used an annulus width of 1$'$ for three faint groups, but a
2$'$ width was used for the bright massive group because of the
small number of massive stars. In addition, we also calculated
the surface density of presumably non-member stars (bright field stars -
stars brighter and redder than the PMS locus; faint field stars - stars 
fainter and bluer than the PMS locus). The error in the surface density 
is assumed to follow Poisson statistics, i.e. $\epsilon = \sqrt{N}/S$ 
(where N and S are the number of stars and the area of the annulus, 
respectively). As expected, the surface density profiles of field stars 
did not show any radial variation. The results are summarized in Table 
\ref{tab_rho}. 

\placetable{tab_rho}

The fifth column of Table \ref{tab_rho} represents the contrast
of surface densities between cluster and field. The contrast is very high
for MS and evolved massive stars, but those of PMS stars are very low.
The surface density of MS/post-MS stars shows a large fluctuation at
$r \gtrsim 10'$ due to the small number of early-type stars in the field.
Although there are large differences in surface density variation
among member groups, the surface density of each group reaches 
the level of the surface density of the field region at $r \approx 10'$
regardless of the stellar mass. From the radial variation of surface density 
profiles, the radius of NGC 6231 was adopted to be 10$'$.

\begin{figure}
\epsscale{1.0}
\plottwo{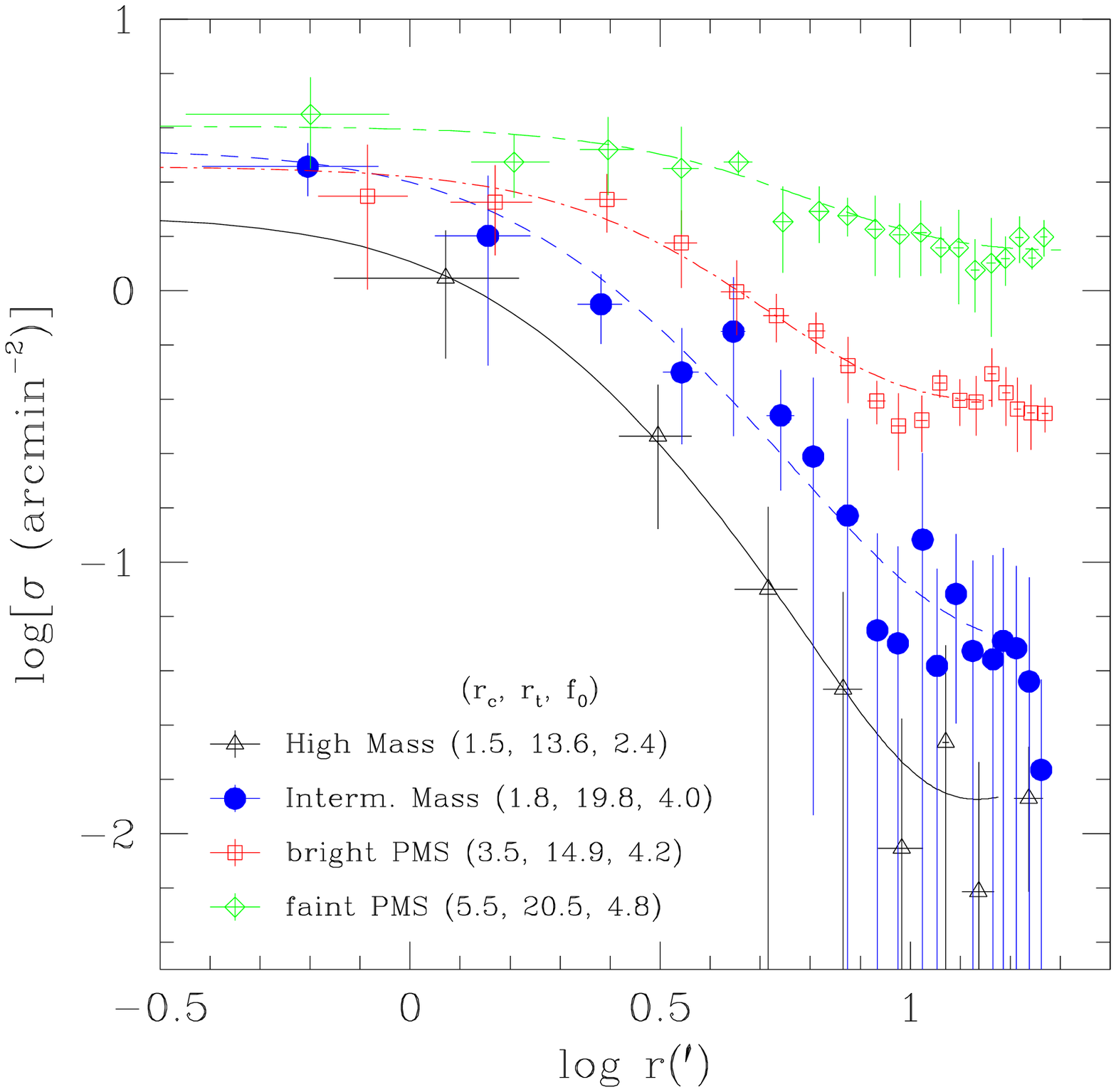}{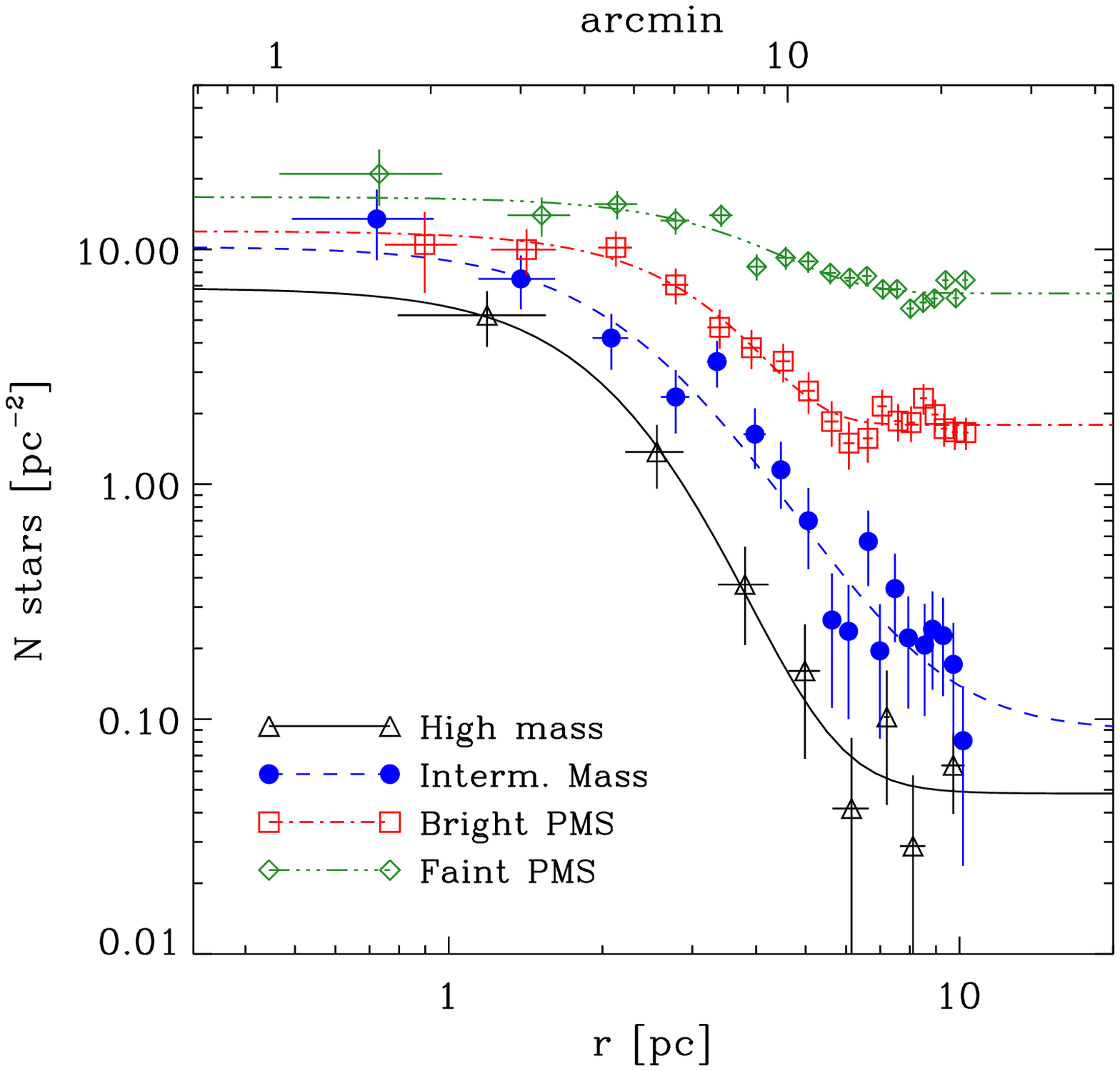}
\caption{Surface density profile fittings. 
(Left) King profile fits. (Right) EFF profile \citep{eff87} fits. 
Note that the two figures use different units.
\label{figfit} }
\end{figure}

\placefigure{figfit}

We also attempted to fit the cluster surface density profiles to a King 
empirical density profile of a spherical system \citep{king62}, as shown in
the left panel of Figure \ref{figfit}. As expected, the tidal radii of all 
groups were very similar to each other, and between 10$'$ -- 20$'$. 
But the core radius varied systematically from 1.$'$5 to 5.$'$5.
As King profiles are strictly speaking not appropriate for open clusters, 
we also fitted the cluster surface density  using \citet{eff87}'s
profiles (EFF profiles). The best fit EFF profile parameters are
presented in Table \ref{tab_rho}. Using the best-fit values,  we computed the core radius of the EFF 
profile, i.e. the radius at which the surface density drops to half 
its center value. As for the King profile, we observed a systematic increase 
from 1.$'$9 to 4.$'$7 (corresponding to 0.9 to 2.2~pc given $V_0 - M_V = 11.0$) for 
populations with decreasing masses.  Both King and EFF profile fittings  
imply a mass segregation within NGC 6231 - massive stars are more 
concentrated in the cluster center, while faint PMS stars are more widely 
distributed. We will come back to this aspect in Section 4.5.

\section{THE INITIAL MASS FUNCTION AND THE AGE OF NGC 6231}

\subsection{Adopted Calibrations and Stellar Evolution Models}

In SBL98 and our studies on other young open clusters, we used 
the spectral type - temperature scale of \citet{C97} for O type stars. 
Advances in astronomical instruments, observing techniques, and non-LTE model 
atmospheres have made an appreciable change in the temperature scale of massive
O type stars. In this study, the spectral type - temperature 
scale of O type stars of \citet{martins05} is adopted. Their calibration
gives somewhat lower temperatures, and hence smaller bolometric corrections
for O type stars than that of \citet{C97}. For intermediate- and 
low-mass stars no significant change in the temperature scale was made,
and therefore the same calibrations were used as in \citet{HSB12}.

In the 2000s, stellar rotation was introduced in the evolutionary models 
of massive stars \citep{mm00}. But in most studies the stellar evolution
models by \citet{schaller92} were used because new models with stellar
rotation were limited to only a few high masses. Recently, new
stellar evolution models with rotation covering a wide mass range were 
published \citep{rot12,brott11}. The stellar evolution models by \citet{brott11}
are limited for the stars with masses 5 -- 60 M$_\odot$. This mass range is
useful for NGC 6231, but our long term project for investigating the shape
of the IMF of young massive clusters (e.g. see \citet{HSB12,lim12}) requires 
a much wide range of stellar masses.
In this study, we will mainly use the stellar evolution models by \citet{rot12}
that include a moderate initial rotation ($v_{eq} = 0.4 v_{crit}$, thus
$\approx 220 km s^{-1}$ for massive main sequence stars).
Although the final ZAMS temperatures and luminosities of a 1 M$_\odot$
star are very similar to each other, the detailed evolutionary tracks
of \citet{bcah98} and \citet{sdf00} differ from each other as
shown in Figure \ref{figcmd}. As the models by \citet{sdf00}
cover a relatively large mass range, we will use the PMS evolution 
models by \citet{sdf00} for the age and mass estimate of low-mass
PMS stars.

\subsection{Hertzsprung-Russell Diagram and Age of NGC 6231}

\begin{figure}
\epsscale{0.5}
\plotone{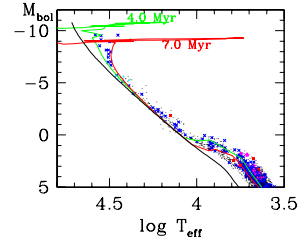}
\caption{The Hertzsprung-Russell diagram of NGC 6231. Triangles represent variables.
The other symbols are the same as Fig. \ref{figccd}. \label{fighrd} }
\end{figure}

\placefigure{fighrd}

The age of an open cluster can be determined by comparing the observations
with theoretical stellar evolution models. The HRD
is the basic diagram to diagnose a stellar system. To construct the HRD
of NGC 6231, we should correct the interstellar reddening of individual stars.
For early-type stars the reddening is estimated from the color-color diagram 
and applied individually. But for low-mass PMS stars, as there is
no direct way to estimate the reddening from photometric data alone, the 
reddening of these stars are estimated from the reddening map presented in
Figure \ref{figebv}. Using the adopted calibrations, we transformed the 
reddening-corrected colors to physical parameters, such as effective 
temperature and bolometric magnitude. For massive O and early B type stars,
the spectral type is the most important indicator of the effective temperature 
and hence bolometric correction as well.

\begin{figure}
\epsscale{0.5}
\plotone{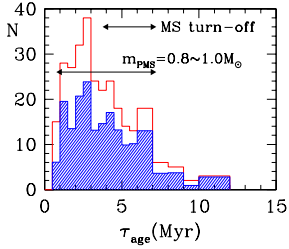}
\caption{The age distribution of stars in NGC 6231. White and hatched histograms
represent, respectively, the age distribution of all the low-mass PMS stars 
within the cluster radius, and those that have been statistically subtracted for 
the contribution of field stars. \label{figsfh} }
\end{figure}

\placefigure{figsfh}

The HRD of NGC 6231 is presented in Figure \ref{fighrd}. The theoretical 
ZAMS and two isochrones of ages 4 Myr and 7 Myr are superimposed. These 
isochrones were derived by interpolating the theoretical stellar evolution 
tracks of \citet{rot12} and the PMS evolution tracks of \citet{sdf00}. 
The age of massive stars in NGC 6231 is between 4 Myr and 7 Myr. This is 
somewhat larger than the age obtained by SBL98 (2.5 -- 4 Myr).
Although there was a change in the stellar evolution models used in 
the age estimate, the main cause of the big difference in age is the change 
in the adopted calibrations, especially for massive O type stars.
For another young open cluster Westerlund 1, the new and old stellar 
evolution models from the Geneva group \citep{rot12,schaller92} do not give 
any significant difference in age \citep{lim12}. However, the stellar evolution models
by \citet{brott11} give a slightly younger age (3 -- 4 Myr) for NGC 6231 
(see section 5.2 for details).

The situation for PMS stars is the opposite - the age of low-mass PMS stars
are younger than presented in SBL98. Many PMS members and candidates 
are slightly redder than the 4 Myr isochrone. The age distribution of
low-mass PMS stars is shown in Figure \ref{figsfh}. The histograms show
the age distribution of all PMS stars in the PMS locus and are corrected for
the contribution of interlopers (Sco OB1 association members and field stars),
respectively. As the PMS evolution models do not give a consistent age
for the whole mass range \citep{sbc04} (see also \citep{sbl97}), the distribution
is derived only for those PMS stars with masses between 0.8 -- 1.0 M$_\odot$. 
The age of low-mass PMS stars is between 1 Myr and 7 Myr (\citet{sana07} also
found a similar spread of age of low-mass PMS stars), and therefore 
the age spread is about 6 Myr, which is the typical spread from CMD analysis
\citep{sb10}. 

\subsection{The Initial Mass Function}

\begin{figure}
\epsscale{0.5}
\plotone{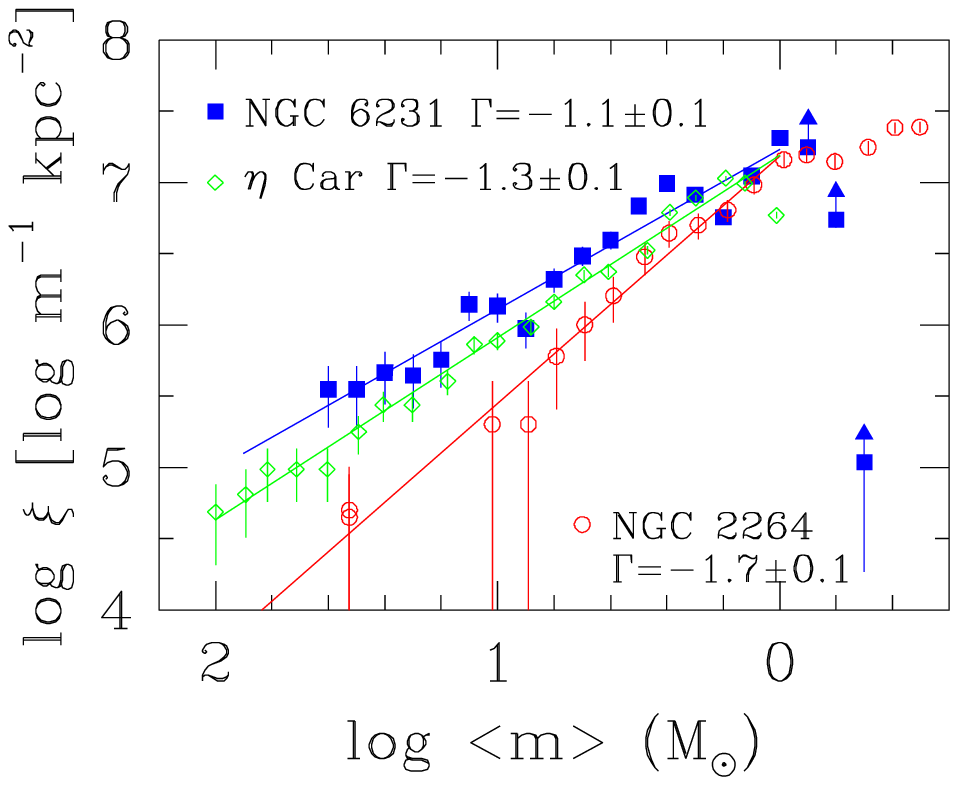}
\caption{The IMF of NGC 6231. Filled squares are the IMF of NGC 6231, 
while open circles and open diamonds are that of NGC 2264 \citep{sb10} 
and the $\eta$ Carina nebula \citep{HSB12}. The solid lines represent the least 
square fit to the IMFs. The slope of the IMF ($\Gamma$) is calculated for 
the stars with $\log m >$ 0.2.
\label{figimf} }
\end{figure}

\placefigure{figimf}

The mass of an individual star is estimated from the HRD. For massive stars
it is not easy, because a small difference in age gives a very different position in
the HRD, and it is therefore impossible to use a single isochrone or 
mass-luminosity relation. In addition, the complex evolutionary tracks of
massive stars make this matter even more difficult. We tried another way to
estimate the mass of massive stars - the mean initial mass of stars at 
a given point in the HRD is calculated from a Monte Carlo simulation
of a stellar system. A few million stars with masses larger than 20 M$_\odot$
were generated for about 500 Myr at a constant rate. The stellar masses
were generated according to an IMF with Salpeter type slope ($\Gamma = -1.3$).
In the model calculation, stars were evolved according to the recent stellar 
evolution models of \citet{rot12}. From the model, we calculated the mean mass
and age of stars within the specified photometric box in the HRD. From
this database we calculated the mass and age of massive evolved stars.

The mass of intermediate-mass MS stars was estimated using the mass-luminosity 
relation of stellar evolution models. 
The mass of PMS stars was estimated by interpolating
the PMS evolution models of \citet{sdf00}. And then the number of stars
in a logarithmic mass interval of $\Delta \log m$ = 0.2 was calculated. The number 
of stars in the field region was also calculated in the same way. After 
subtracting the contribution of field stars, the cluster mass function was 
then obtained by dividing the projected area and mass interval. To minimize 
the effect of binning, we also calculated the mass function by shifting
the mass bin by $\log m$ = 0.1. Actually we only estimated the initial mass
of individual stars, not the current mass of the star. Therefore the mass
function is the same as the IMF except for the most massive bin for
the clusters with age $\geq$ 3.5 Myr. For NGC 6231, the mass function obtained
here is the IMF of NGC 6231 for $m \lesssim$ 45 M$_\odot$.

The IMF of NGC 6231 is presented in Figure \ref{figimf}. The IMFs of NGC 2264
\citep{sb10} and the $\eta$ Carina nebula \citet{HSB12} are also presented.
The slope of the IMF of NGC 6231 is slightly shallower than that of the $\eta$ 
Carina nebula, but the difference is marginal because the shallow slope is 
partly caused by the dearth of stars at $\log m \approx$ 0.2. But the difference
between NGC 6231 and NGC 2264 is evident. The slope of NGC 2264 is very steep.
Although the surface density at $\log m \approx$ 0 is very similar, the surface
density at $\log m \approx 1$ of NGC 6231 is about 10 times higher than that 
of NGC 2264. Recently \citet{damiani09} obtained a slightly steep IMF ($\Gamma 
=$ -1.3 -- -1.5) for NGC 6231. Such a difference in the IMF slope may be
caused by the difference in the adopted distance, the amount of reddening
correction, and/or the adopted stellar/PMS evolution models.

\subsection{The Total Mass of NGC 6231}

The total mass of a cluster is an important parameter. Recently 
\citet{weidner10} studied the relation between the most massive star and its
parental cluster mass. They classified clusters into three categories
according to their total cluster mass - lowest mass clusters ($M_{cl} \leq 10^2$
M$_\odot$), moderate mass clusters ($10^2$ M$_\odot < M_{cl} \leq 10^3$ 
M$_\odot$), and rich clusters ($M_{cl} > 10^3$ M$_\odot$). And they found a plateau
of a constant maximal star mass ($m_{max} \approx 25$ M$_\odot$) for clusters
with masses between $10^3$ M$_\odot$ and $4 \times 10^3$ M$_\odot$. 

We calculated the total mass of NGC 6231 by simulating model clusters
using a Monte Carlo method. For the model clusters we assumed the IMF
of NGC 6231 in four cases for the low-mass regime - (1) the IMF of NGC 6231 and
that of NGC 2264 \citep{sb10} were adjusted between $\log m =$ 0.6 -- 0.4, and we 
used the adjusted IMF of NGC 2264 for $\log m <$ 0.4, (2) the IMF of NGC 6231 
and that of NGC 2264 were adjusted between $\log m =$ 0.2 -- 0.0, and 
we used the adjusted IMF of NGC 2264 for $\log m <$ 0.0, (3) the IMF of NGC
2264 was used for $\log m <$ 0.0 without any adjustment, and (4) the IMF of
\citet{kroupa02}. For the massive part of the first three cases,
we simply extrapolated the IMF of NGC 6231 up to $\log m =$ 2.0. We assumed
the upper mass limit of stellar mass as 100 M$_\odot$, and an age spread of
about 3 Myr (age: 4 -- 7 Myr) for the massive stars ($m \geq$ 30 M$_\odot$) 
as obtained in section 4.2. A total of 100,000 stars
were generated, and a scaling factor was calculated to
reproduce the number of stars with masses between $\log m =$ 0.4 -- 1.0. 
The total masses estimated were $3.0 \times 10^3$M$_\odot$, $2.2 \times 
10^3$ M$_\odot$, $2.1 \times 10^3$ M$_\odot$, and $3.3 \times 10^3$ M$_\odot$, 
respectively. And therefore the total mass of NGC 6231 is 2.6 ($\pm$ 0.6) 
$\times 10^3$ M$_\odot$. This value is slightly smaller than the total 
mass of NGC 6231 ($4595^{-2312}_{+4676}$ M$_\odot$) estimated by 
\citet{weidner10}. Such a large discrepancy in the total mass may be related 
to the subtraction of field contamination, especially for stars in the Sco 
OB1 association.

\subsection{MASS SEGREGATION} 

\begin{figure}
\epsscale{0.9}
\plotone{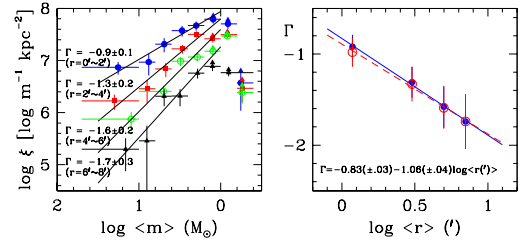}
\caption{Radial variation of the IMF. (Left) The IMF at different radii.
(Right) The radial variation of the slope of the IMF. The large dots, the solid line,
and the slope gamma represent the radial variation for the case of no correction for
the data incompleteness, while open circles and a dashed line are corrected 
for the data incompleteness. The slope of the IMF varies systematically
from cluster center to the outer region. The IMF at the outermost region is
not shown because of its large uncertainty due to the small number of cluster
stars.
\label{figvimf} }
\end{figure}

Mass segregation in young open clusters is a hot issue in the 
dynamical evolution of star clusters. \citet{mvpz07} proposed a plausible 
scenario for mass segregation as mentioned in the introduction. NGC 6231 is 
an important target for such a study as the cluster is little obscured, 
has many members, and relatively small differential reddening across 
the cluster. To see the extent of mass segregation in NGC 6231, 
we calculated the radial variation of the IMF as shown in Figure \ref{figvimf}.
The IMF for each ring (width of the ring = 2$'$) was calculated in the same 
way as in section 4.3. But as the number of massive stars was very small, 
only one mass bin was used for the stars more massive than 10 M$_\odot$. 
For the outermost ring ($r = 8' \sim 10'$), the number of stars in the ring,
after subtracting the field contribution, was negative in many cases, and
therefore we did not try to calculate the IMF for the ring. The completeness
of photometry was better than 90\% even in the innnermost ring for $\log m >$
0.4, so we did not correct for the effect of data incompleteness in the left 
panel of Figure \ref{figvimf}. But the effect of incompleteness in $\log m < 
0.4$ cannot be neglected, especially for the innermost ring where the crowding
effect is most severe. In the right panel of Figure \ref{figvimf}, we also calculated 
and presented the slopes of the IMF 
for a given ring, with and without correction for data incompleteness.

\placefigure{figvimf}

As can be seen in Figure \ref{figvimf}, the slope of the IMF increases
systematically from the center to the peripheral region. The slope of the IMF 
in each ring was calculated, and shows a variation with radius from  the cluster
center (right panel). The slope of the IMF varies
systematically - it is very shallow in the center and very steep at the peripheral 
regions. The right panel of Figure \ref{figvimf} shows that the mass segregation
in NGC 6231 is evident and very systematic.

\begin{figure}
\epsscale{0.5}
\plotone{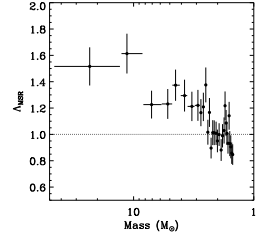}
\caption{The relation between the mass segregation ratio ($\Lambda_{\rm MSR}$) and 
the stellar mass. The number of member stars in each bin is 20. The dot represents
the average mass of each bin, and the horizontal bar represents the range
of stellar masses for a given bin.
The $\Lambda_{\rm MSR}$ of the stars with $m \gtrsim 3 M_\odot$ shows
mass segregation, while those of low-mass stars have about unity 
($\Lambda_{\rm MSR} \approx 1$ - no mass segregation).
\label{figmst} }
\end{figure}

\placefigure{figmst}

\citet{rja09a} introduced a parameter, the mass segregation ratio $\Lambda_{MSR}$,
to detect and quantify the level of mass segregation in a cluster. The
$\Lambda_{MSR}$ compares the path length from the minimum spanning tree of 
a certain kind of objects and that of random samples. The merit of this
method is that the $\Lambda_{MSR}$ is insensitive to uncertainty in determining
the center of an open cluster. \citet{sana10} applied
this method to the young compact cluster in the $\eta$ Carina nebula, Trumpler 
14, and found a signature of mass segregation for bright members ($m \gtrsim
10$ M$_\odot$), and marginal
mass segregation for less bright members ($K_s \lesssim$ 10.5 mag). We also
calculated the mass segregation ratios for member stars (MS stars and PMS 
stars in the PMS locus within the adopted cluster radius) and present it in
Figure \ref{figmst}. In the calculation of $\Lambda_{MSR}$, we used the same 
bin size (N = 20). The most massive two bins ($m \gtrsim 8$ M$_\odot$) show a distinct 
mass segregation, and stars with masses between 10 -- 3 M$_\odot$ are 
marginally mass segregated. It is worth noting that as the completeness of 
photometry is lower for the fainter stars, especially fainter stars in 
the center, the $\Lambda_{MSR}$ is affected by the data incompleteness. 
Given the limitation, we can say that massive stars ($m \gtrsim$ 8 M$_\odot$) 
in NGC 6231 are mass segregated. Discussion on the origin of mass segregation 
is presented in section 5.3.

\section{DISCUSSION}

\subsection{HD 153919}

The runaway O6.5Iaf star HD 153919 has been proposed to have originated from
Sco OB1 \citep{msb74,ankay01}.
If the location of HD 153919 ($\mu_\alpha = 2.80 \pm 0.59 mas/yr,
~ \mu_\delta = 4.71 \pm 0.32 mas/yr$, \citet{leeuwen07}) before the supernova 
explosion is assumed to be at the center of the distribution of polarization vectors \citep{fmmvbv}, we can check 
whether or not that is a reasonable assumption, on the basis of stellar evolution theory. 
As the angular distance between the the polarization center and HD 153919 is 
($\Delta \alpha = 1.^\circ 92, ~ \Delta \delta = 3.^\circ84, ~\rho = 
4.^\circ29$), the proper motion of the star should have the minimum value ($\mu - 
\epsilon_\mu$) for the star to have crossed NGC 6231 in the past. And therefore if
we take the proper motion of the star as $\mu_\alpha' = 2.21 mas/yr,
~ \mu_\delta' = 4.39 mas/yr$, then the star would have been ejected from the polarization center
3.14 $\pm$ 0.01 Myr ago. As the age of massive stars in NGC 6231 is
4 -- 7 Myr, the past supernova exploded when the age of NGC 6231 was
1 -- 4 Myr. 

The life-time of the most massive star (120 M$_\odot$)
in \citet{rot12} is 3.5 Myr, and that of a 85 M$_\odot$ is 4.0 Myr.
If HD 153919 originated from NGC 6231, the initial mass of the supernova
progenitor would be 85 M$_\odot$, or larger. This result is consistent with
the current stellar evolution models and the IMF of NGC 6231, as we could
expect 2 -- 3 stars more massive than the most massive star in the
current NGC 6231. 
While the end product of the evolution of star typically more massive than 25 
M$_\odot$ is expected to be a black hole, a massive star may end its life 
as a neutron stars, if it loses enough mass through wind mass-loss 
\citep{fryer02} and/or binary interaction with a lower mass companion.
As most massive stars are in close binary systems \citep{sana08}, part of the
primary's mass is likely to be transferred to the companion during the MS stage or in 
later evolutionary stages, and finally it becomes a neutron star after a supernova explosion. 
Given that HD153919 is a high-mass X-ray binary, the second scenario is the more likely.

\subsection{Age of Massive Stars}

\begin{figure}
\epsscale{1.0}
\plottwo{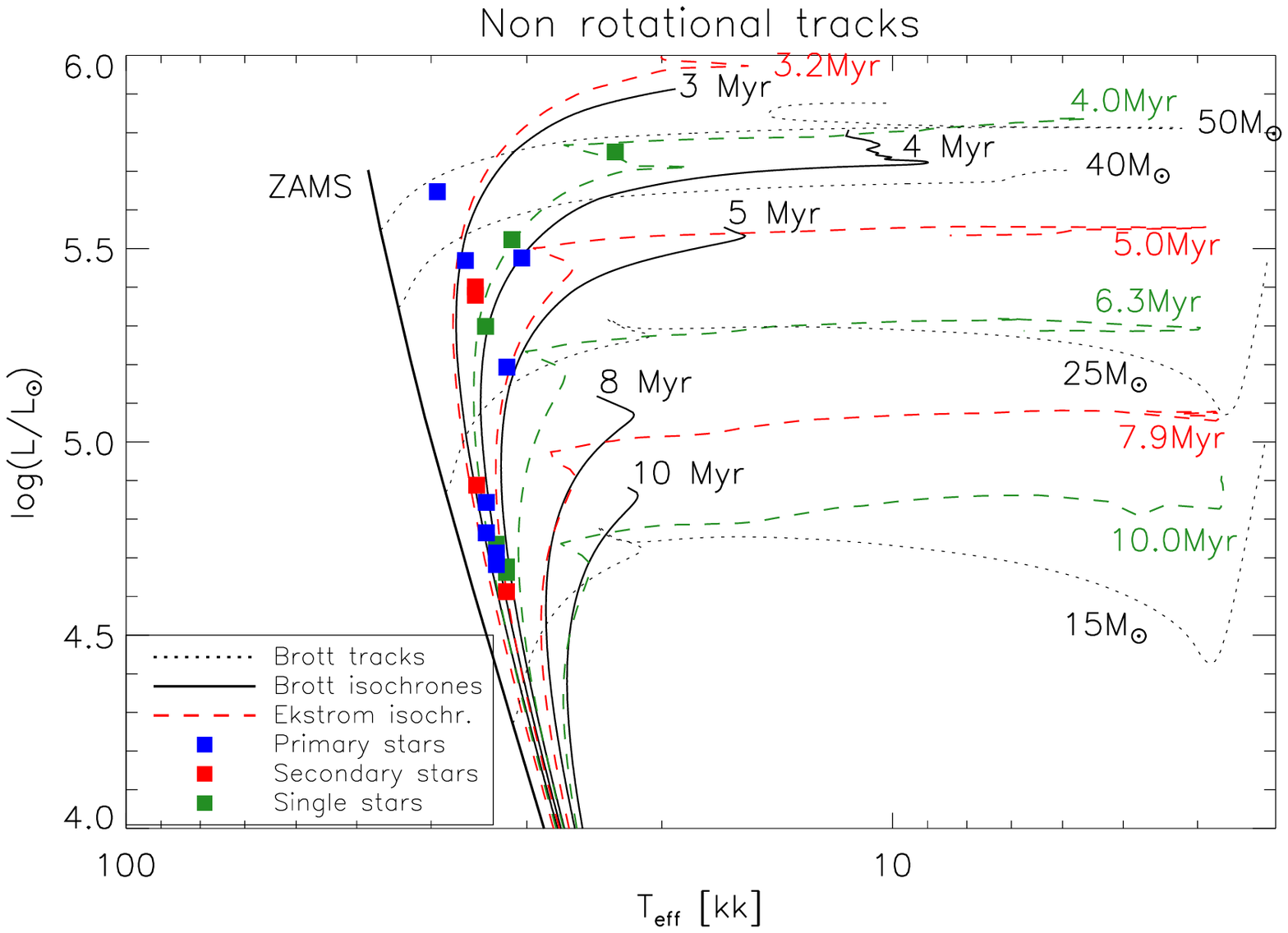}{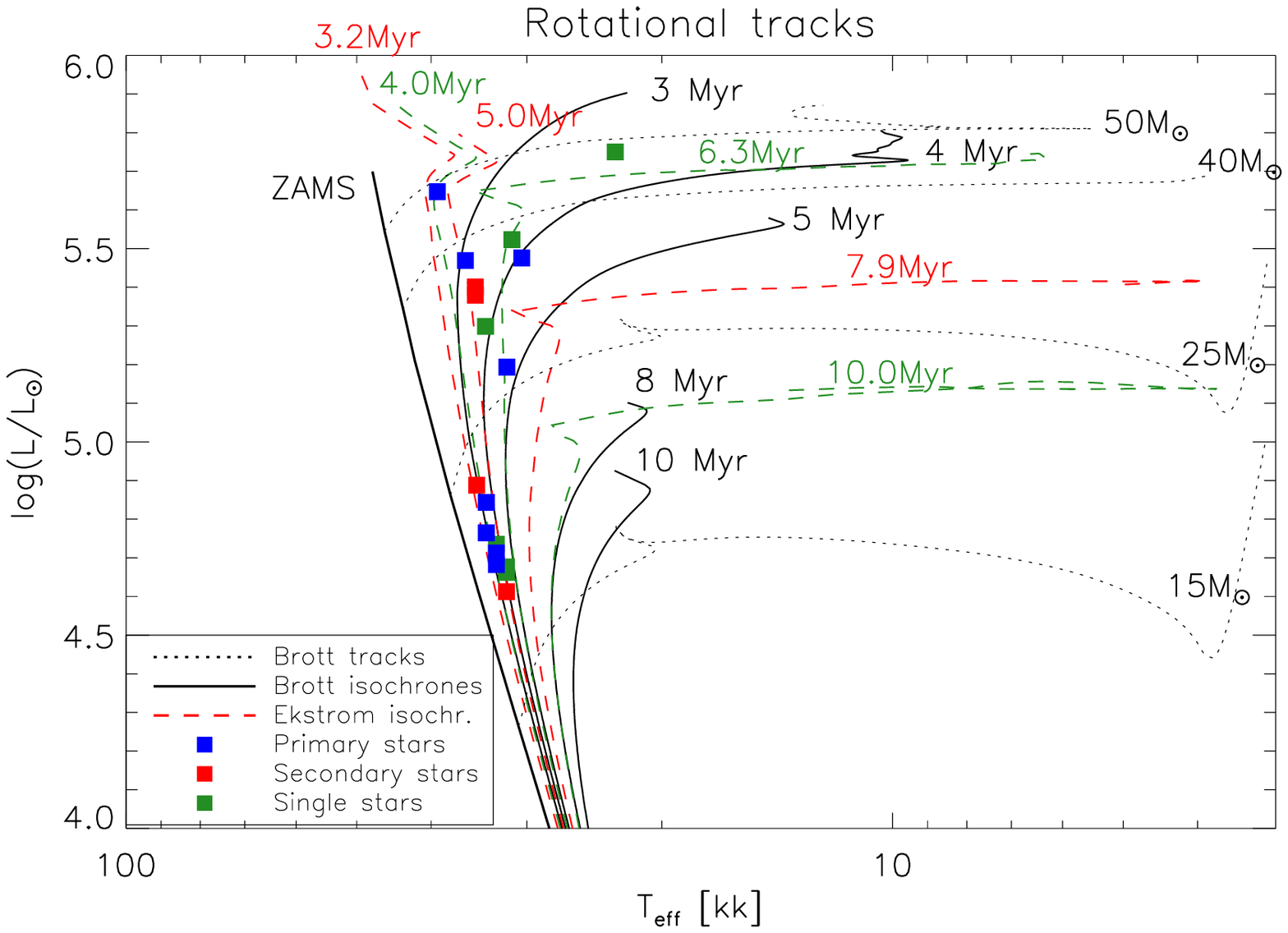}
\caption{The upper part of the Hertzsprung-Russell diagram.
(Left) Non-rotating models, (Right) rotating models.
\label{figuhrd}}
\end{figure}

\placefigure{figuhrd}

The current state-of-the-art stellar evolution models do include the effect 
of stellar rotation. 
%Two sets of models are currently available for high mass 
%stars: those from the Bonn \citep{brott11} and Geneva (with the latest update 
%by \citet{rot12}) teams. 
It is beyond the scope of this discussion to provide 
a comparison between the different assumptions as well as the implementation 
of different physical mechanisms; however, we tested the impact of adopting 
the different available models on the derived properties of the high-mass stars, 
and more specifically on their derived age.

Figure \ref{figuhrd} (left panel) compares the isochrones obtained from 
the non-rotating evolutionary models of Brott et al. and Ekstrom et al. 
The main difference resides in the position of the end of the main sequence 
in the HRD and  results from different assumptions on the value of 
the core overshooting parameter. The age derived for the massive stars is, however,
very similar, and a single isochrone, corresponding to an age of about 
$3.5 \pm 0.5$ Myr, reproduces most of the single and binary O star population 
in NGC 6231.

The age obtained from evolutionary tracks that incorporate initial stellar 
rotation is, however, somewhat different in both sets of models. The \citet{rot12}
rotating models adopt an initial $v_\mathrm{eq}$ of $0.4 v_\mathrm{crit}$, 
which is of the order of 200 - 220 km s$^{-1}$ for the mass range considered. 
Figure \ref{figuhrd} (right panel) compares the Ekstrom et al. and  
Brott et al. models with similar initial rotation rates.
For the Brott et al. models, the location of the isochrones is little affected and 
our previous age estimates stand. However, the Ekstrom et al. rotating 
isochrones are shifted to higher temperatures so older isochrones, 
of 4 to 7 Myr, are needed to reproduce the HRD position of the NGC 6231 O-type 
stars. One has to adopt an initial rotation larger than 400~km/s for the Brott 
et al. models to fit a similar cluster age. Such a large initial rotation 
is unlikely, however, given that the majority of unevolved O-type stars are 
relatively slow rotators (e.g., \citep{penny09}).

The question of whether the formation of high-mass stars in a star forming 
region triggers or suppresses the formation of lower mass stars (see e.g.
\citet{herbig62}) can in principle 
be tested by comparing the age of the high- and low-mass stellar population in 
NGC 6231. Unfortunately, uncertainties in the evolutionary calibrations makes 
it difficult to age-date the high-mass stars with sufficient confidence.
 
If the bulk of the NGC 6231 high mass stars are truly 5 or 6 Myr old, they would 
be born at the very beginning of the star forming event that lead to the formation 
of NGC 6231. On the other hand, an age of 3.5 $\pm$ 0.5 Myr old would see
them formed when star formation activity was at its peak (see 
Fig. \ref{figsfh}).

\subsection{Primordial Mass Segregation Or Dynamical Evolution}

The current theoretical point of view on mass segregation in young open 
clusters is that it results from the rapid dynamical evolution of clumps in 
the subvirial state \citep{mvpz07,rja09b}. This picture implies that 
there should be several clumps in a star forming region, and that
the star forming region may be, on average, less dense than the case of only 
a single star forming clump in a star forming region. This means
that massive stars could be formed in the peripheral region of a star forming
region. In this case,
we could have a chance to observe the formation process of massive stars with
masses over 30 M$_\odot$, which would be extremely important, as we still we 
have no detailed information on the formation processes of massive stars.

The surface density profiles and the radial variation of the IMF
show clearly that NGC 6231 is mass-segregated. And the age of NGC 6231
is much older than the time required for mass segregation in a subvirial
cluster with substructure (see Fig. 2 of \citet{rja09b}). So, is the mass segregation
in NGC 6231 of dynamical origin? Other types of data can help decide. 
The binary fraction of massive O type stars in NGC 6231 \citep{sana08} 
is 100\% at the center, and it reaches about 60 \% from 2$'$ up to 15$'$
(Sco OB1 region). The mass of a binary system is more massive than single 
stars, but the binary fraction of O type stars is nearly the same, even in the 
peripheral region of NGC 6231. This fact implies that the mass segregation 
in NGC 6231 may be of primordial origin rather than the  dynamical 
evolution scenario of \citet{mvpz07,rja09b}.

\section{SUMMARY}

New wide-field photometry for the young open cluster NGC 6231 is presented
and analyzed. The main results obtained from this study are summarized 
as follows.

(1) The reddening law toward NGC 6231 was derived from color excess ratios
from the optical to mid-IR {\it Spitzer} IRAC bands. The reddening law was 
slightly higher ($R_V = 3.2$), but very close to the normal reddening law.
Interestingly, stars in the cluster center show relatively larger color
excess ratios.  This fact, and the relatively smaller reddening in the center, are
conjectured as the result of strong stellar winds from massive stars or a
supernova that exploded a few Myr ago.

(2) The IMF and age of NGC 6231 were derived. The age of NGC 6231 is 
4 -- 7 Myr ($\Delta \tau_{\rm MS}$ = 3 Myr) for massive stars, and 1 -- 7 Myr 
($\Delta \tau_{\rm PMS}$ = 6 Myr) for low-mass
PMS stars using a new stellar evolution models with rotation \citep{rot12}
and by adopting new spectral type - temperature relation of O type stars 
\citep{martins05}. The slope of the IMF is slightly shallow, but very close
to the Salpeter value ($\Gamma$ = -1.1 $\pm$ 0.1). In addition, the total mass
of NGC 6231 is estimated to be about $2.6 (\pm 0.6) \times 10^3$ M$_\odot$ from simulated
model clusters using a Monte Carlo method.

(3) The surface density profiles of four mass groups were derived. Although 
there was a large difference in the ratio of central density to field density, 
the four groups showed similar cluster radii. In addition, the radial variation
of the mass function was derived and showed an evident mass segregation
in NGC 6231. The mass segregation of massive stars ($m \gtrsim$ 8 M$_\odot$)
was also confirmed from the higher value of the mass segregation ratio $\Lambda_{MSR}$
from the minimum spanning tree. The mass segregation in NGC 6231 may be of 
primordial origin and not as a result of dynamical evolution, because of 
the constancy of the binary fraction of massive O type stars with radius.

(4) The runaway O type supergiant HD 153919 could have originated from NGC 
6231. Using the {\it Hipparcos} proper motion data, the past supernova,
if it exploded at the center of the polarization \citep{fmmvbv},
would have exploded about 3.1 Myr ago, and the initial mass of the supernova
progenitor is estimated to be 85 M$_\odot$ or higher.

\acknowledgments 
The authors thank the anonymous referee for many insightful comments.
H.Sung acknowledges the support of the National Research Foundation of Korea (NRF)
funded by the Korea Government (MEST) (Grant No.20120005318). H.Sung would like
to express his thanks to Professor Harvey Butcher, the director of Research 
School of Astronomy and Astrophysics, ANU for hosting him as a Visiting Fellow.

\clearpage

\clearpage

\begin{deluxetable}{llcccc}
\tablecolumns{6}
\tabletypesize{\scriptsize}
\tablecaption{Standard Stars, Atmospheric Extinction Coefficients, time variation coefficients, and photometric zero points \label{tab_coef} }
\tablewidth{0pt}
\tablehead{
\colhead{Date of Obs.} & \colhead{standard stars} & \colhead{Filter} & \colhead{Extinction Coefficients} & \colhead{Time Variation} & \colhead{Photometric Zero Points\tablenotemark{a}} }

\startdata
               & & $I$ & 0.101 $\pm$ 0.007 &         0.000        & 22.285 $\pm$ 0.011 \\
               & E5-R\tablenotemark{c}, E5-O\tablenotemark{c}, E5-T\tablenotemark{c} & $V$ & 0.182 $\pm$ 0.004 & -0.0014 $\pm$ 0.0008 & 22.791 $\pm$ 0.009 \\
2000, June, 22 & E6-U, E661, E6-Y, E6-W & $B$ & 0.294 $\pm$ 0.007 & -0.0035 $\pm$ 0.0016 & 22.640 $\pm$ 0.011 \\
               & E7-W, E737, E7-c, E7-X & $U$ & 0.533 $\pm$ 0.016 & -0.0056 $\pm$ 0.0034 & 20.213 $\pm$ 0.022 \\
               & & H$\alpha$ & 0.101 $\pm$ 0.017 & \nodata        & 18.067 $\pm$ 0.024 \\ \hline
               & & $I$ & 0.070 $\pm$ 0.006 & +0.0063\tablenotemark{b} $\pm$ 0.0011 & 22.284 $\pm$ 0.008 \\
               & E5-R\tablenotemark{c}, E5-O\tablenotemark{c}, E5-T\tablenotemark{c} & $V$ & 0.134 $\pm$ 0.005 & -0.0048\tablenotemark{b} $\pm$ 0.0014 & 22.760 $\pm$ 0.008 \\
2000, June, 25 & E6-U, E661, E6-Y, E6-W & $B$ & 0.271 $\pm$ 0.006 & -0.0041\tablenotemark{b} $\pm$ 0.0014 & 22.622 $\pm$ 0.009 \\
               & E7-W\tablenotemark{c}, E737\tablenotemark{c}, E7-c\tablenotemark{c}, E7-X\tablenotemark{c} & $U$ & 0.506 $\pm$ 0.012 & -0.0044\tablenotemark{b} $\pm$ 0.0014 & 20.193 $\pm$ 0.018 \\
               & & H$\alpha$ & 0.092 $\pm$ 0.010 & \nodata        & 18.072 $\pm$ 0.003 \\
\enddata

\tablenotetext{a}{Photometric zero points at midnight}
\tablenotetext{b}{for UT $\leq$ 13$^h$}
\tablenotetext{c}{observed twice - once at meridian, the other at air mass 
$\approx$ 2.}

\end{deluxetable}

%\begin{deluxetable}{rcccc}
%\tablecolumns{5}
%\tabletypesize{\scriptsize}
%\tablecaption{Standard Stars \label{tab_std}}
%\tablewidth{0pt}
%\tablehead{
%\colhead{Date of Obs.} & \colhead{standard star} & \colhead{No. of Obs.} & 
%

\begin{deluxetable}{rcccccccccccccccccccl}
\tablecolumns{21}
\tabletypesize{\scriptsize}
\rotate
\tablecaption{Photometric Data\tablenotemark{a} \label{tab_data}}
\tablewidth{0pt}
\tablehead{
\colhead{ID} & \colhead{$\alpha_{\rm J2000}$} & \colhead{$\delta_{\rm J2000}$} & \colhead{$V$} &
\colhead{$V-I$} & \colhead{$B-V$} & \colhead{$U-B$} &
\colhead{$\epsilon_{V}$} & \colhead{$\epsilon_{V-I}$} & \colhead{$\epsilon_{B-V}$} &
\colhead{$\epsilon_{U-B}$} & \multicolumn{4}{c}{N$_{\rm obs}$} & \colhead{2MASS} &
\colhead{M\tablenotemark{b}} & \colhead{D/V\tablenotemark{c}} & \colhead{SBL98}  &
\colhead{Sp} & \colhead{remark} }

\startdata
  3610  &  16:54:17.95 &  -41:48:01.6 &  11.063 &   0.355 &   0.210 &  -0.492 &   0.001 &   0.002 &   0.008 &   0.005 &  2 &  2 &  2 &  2 & 16541793-4148013 &   &   &    480 & B2IVn   & Se213               \\
  3611  &  16:54:17.96 &  -41:44:57.1 &  16.460 &   1.256 &   1.060 &   0.654 &   0.000 &   0.002 &   0.004 &   0.021 &  4 &  4 &  3 &  2 & 16541795-4144568 &   &   &   6777 &         &                     \\
  3612  &  16:54:17.97 &  -42:00:37.3 &  17.808 &   1.717 &   1.345 &   0.809 &   0.013 &   0.027 &   0.026 &   0.108 &  1 &  1 &  1 &  1 & 16541794-4200373 &   &   &        &         &                     \\
  3613  &  16:54:17.98 &  -42:09:07.4 &  12.898 &   0.843 &   0.650 &   0.129 &   0.000 &   0.003 &   0.001 &   0.003 &  3 &  3 &  3 &  4 &                  &   &   &        &         &                     \\
  3614  &  16:54:17.98 &  -41:31:22.8 &  16.849 &   1.311 &   1.049 &   0.496 &   0.019 &   0.020 &   0.022 &   0.036 &  2 &  2 &  1 &  1 & 16541799-4131229 &   &   &        &         &                     \\
  3615  &  16:54:17.99 &  -41:56:12.5 &  17.182 &   1.592 &   1.250 &   0.651 &   0.001 &   0.002 &   0.017 &   0.060 &  2 &  2 &  1 &  1 & 16541798-4156125 &   &   &   6779 &         &                     \\
  3616  &  16:54:18.01 &  -41:32:47.3 &  17.420 &   2.991 &   2.393 & \nodata &   0.005 &   0.014 &   0.054 & \nodata &  2 &  2 &  1 &  0 & 16541801-4132474 &   &   &        &         &                     \\
  3617  &  16:54:18.01 &  -41:59:12.7 &  17.577 &   1.667 &   1.334 &   0.873 &   0.022 &   0.029 &   0.027 &   0.081 &  2 &  2 &  1 &  1 & 16541798-4159126 &   &   &   6781 &         &                     \\
  3618  &  16:54:18.03 &  -41:45:11.8 &  14.154 &   0.907 &   0.670 &   0.381 &   0.007 &   0.006 &   0.005 &   0.005 &  4 &  4 &  4 &  3 & 16541802-4145117 &   &   &    481 &         & Se188               \\
  3619  &  16:54:18.17 &  -41:48:13.0 &  17.542 &   1.414 &   1.273 & \nodata &   0.000 &   0.011 &   0.030 & \nodata &  2 &  2 &  1 &  0 & 16541819-4148125 &   &   &   6790 &         &                     \\
  3620  &  16:54:18.17 &  -41:50:16.9 &  13.065 &   0.572 &   0.370 &  -0.040 &   0.002 &   0.007 &   0.012 &   0.000 &  3 &  3 &  3 &  3 & 16541814-4150164 &   &   &    482 & B9V     & Se225               \\
  3621  &  16:54:18.18 &  -41:53:36.9 &  13.337 &   0.608 &   0.420 &   0.105 &   0.011 &   0.012 &   0.011 &   0.003 &  2 &  2 &  2 &  2 & 16541819-4153372 &   &   &    483 &         &                     \\
  3622  &  16:54:18.19 &  -41:45:33.1 &  12.094 &   2.350 &   1.865 &   2.328 &   0.001 &   0.017 &   0.006 &   0.011 &  3 &  2 &  3 &  2 & 16541820-4145330 &   &   &    484 &         &                     \\
  3623  &  16:54:18.22 &  -41:33:59.2 &  16.877 &   1.218 &   0.971 &   0.470 &   0.005 &   0.011 &   0.009 &   0.045 &  2 &  2 &  2 &  1 & 16541823-4133592 &   &   &        &         &                     \\
  3624  &  16:54:18.27 &  -41:34:12.0 &  13.276 &   0.568 &   0.384 &   0.229 &   0.003 &   0.005 &   0.005 &   0.008 &  2 &  2 &  2 &  2 & 16541828-4134120 &   &   &        &         &                     \\
  3625  &  16:54:18.29 &  -41:50:10.1 &  17.168 &   1.827 & \nodata & \nodata &   0.061 &   0.052 & \nodata & \nodata &  2 &  2 &  0 &  0 & 16541824-4150097 &   &   &   6792 &         &                     \\
  3626  &  16:54:18.29 &  -41:51:35.2 &   9.597 &   0.360 &   0.181 &  -0.684 &   0.007 &   0.012 &   0.009 &   0.007 &  1 &  1 &  1 &  1 & 16541832-4151357 &   & V &    486 & B1V(n)  & Se238,HD 326330, bCep\\
  3627  &  16:54:18.30 &  -41:46:38.3 &  14.640 &   1.855 &   1.629 &   1.696 &   0.015 &   0.002 &   0.003 &   0.019 &  4 &  4 &  4 &  2 & 16541828-4146381 &   &   &    485 &         & Se204               \\
  3628  &  16:54:18.33 &  -41:32:23.7 &   8.454 &   0.234 &   0.116 &  -0.456 &   0.016 &   0.038 &   0.024 &   0.018 &  1 &  1 &  1 &  1 & 16541837-4132240 &   &   &        & B9Ib-II & HD 152269           \\
  3629  &  16:54:18.34 &  -41:45:00.0 &  17.447 &   1.646 &   1.258 &   0.711 &   0.022 &   0.001 &   0.005 &   0.147 &  3 &  3 &  2 &  2 & 16541832-4144599 &   &   &   6800 &         &                     \\
  3630  &  16:54:18.37 &  -41:35:23.0 &  17.462 &   1.440 &   1.175 &   0.918 &   0.008 &   0.050 &   0.013 &   0.082 &  2 &  2 &  1 &  1 & 16541836-4135228 &   &   &        &         &                     \\
\enddata

\tablenotetext{a}{Table \ref{tab_data} is presented in its entirety in
the electronic edition of the Astronomical Journal. A portion is shown here
for guidance regarding its form and content. Units of right ascension are
hours, minutes, and seconds of time, and units of declination are degrees,
arcminutes, and arcseconds.}
\tablenotetext{b}{membership - X: X-ray emission star, H: H$\alpha$ emission star, h: H$\alpha$ emission candidate, +: X + H, -: X + h}
\tablenotetext{c}{duplicity or variability - D: stars whose PSF shows a double, but measures as a single star, V: variable from the literature, variable types if available noted in the last column (dSct: $\delta$ Scuti type variable, bCep: $\beta$ Cephei type variable)}
\end{deluxetable}

\begin{deluxetable}{lcccccccccc}
\tablecolumns{9}
\tabletypesize{\scriptsize}
\tablecaption{Comparison with Photoelectric and CCD Photometry \label{tab_comp}}
\tablewidth{0pt}
\tablehead{
\colhead{Author} & \colhead{$\Delta V$} & \colhead{n(n$_{\rm ex}$)\tablenotemark{b}} & 
\colhead{$\Delta (V-I)$} & \colhead{n(n$_{\rm ex}$)\tablenotemark{b}} & \colhead{$\Delta (B-V)$} &
\colhead{n(n$_{\rm ex}$)\tablenotemark{b}} & \colhead{$\Delta (U-B)$} & \colhead{n(n$_{\rm ex}$)\tablenotemark{b}} }

\startdata
\citet{bbg66}      & -0.022 $\pm$ 0.053 & 10 (2) & \nodata & \nodata & +0.009 $\pm$ 0.023 & 12 (0) & -0.027 $\pm$ 0.045 & 11 (1) \\
\citet{ff68}       & +0.025 $\pm$ 0.132 & 7 (1) & \nodata & \nodata & +0.017 $\pm$ 0.041 & 7 (1) & -0.054 $\pm$ 0.061 & 7 (1) \\
\citet{ws68}       & -0.017 $\pm$ 0.051 & 24 (1) & \nodata & \nodata & +0.007 $\pm$ 0.068 & 23 (1) & +0.012 $\pm$ 0.091 & 23 (1) \\
\citet{shs69}       & -0.024 $\pm$ 0.017 & 16 (1) & \nodata & \nodata & -0.003 $\pm$ 0.014 & 16 (1) & -0.003 $\pm$ 0.030 & 17 (0) \\
\citet{gs79}       & -0.007 $\pm$ 0.017 & 16 (1) & \nodata & \nodata & +0.000 $\pm$ 0.018 & 23 (0) & +0.026 $\pm$ 0.040 & 23 (0) \\
\citet{hw84}       & +0.010 $\pm$ 0.069 & 9 (0) & \nodata & \nodata & -0.017 $\pm$ 0.040 & 9 (0) & -0.005 $\pm$ 0.040 & 8 (1) \\
\citet{phyb90}     & +0.017 $\pm$ 0.036 & 24 (4) & \nodata & \nodata & +0.013 $\pm$ 0.015 & 26 (1) & -0.009 $\pm$ 0.035 & 26 (1) \\
\citet{sbl98}\tablenotemark{a} & -0.016 $\pm$ 0.021 &1157(148)& +0.027 $\pm$ 0.023 &1227(74)& +0.012 $\pm$ 0.018 &1114(173)& +0.033 $\pm$ 0.065 &998(122) \\
\citet{pcb98}      & +0.059 $\pm$ 0.056 & 58 (9)& +0.078 $\pm$ 0.064 & 61 (6) &  \nodata          &\nodata &       \nodata      & \nodata\\
\citet{bvf99}\tablenotemark{a} & +0.010 $\pm$ 0.048 & 466(102)& +0.056 $\pm$ 0.050 & 493(50)& +0.000 $\pm$ 0.055 & 496(62)& +0.018 $\pm$ 0.055 &271(60) \\
\enddata

\tablenotetext{a}{Comparison made for $V <$ 16 for $U-B$, and $V <$ 17 for $V$, $V-I$, and $B-V$.}
\tablenotetext{b}{The number of stars in parenthesis is excluded in the comparison}

\end{deluxetable}

\begin{deluxetable}{lcccccc}
\tablecolumns{7}
\tabletypesize{\scriptsize}
\tablecaption{Summary of Surface Density Analysis \label{tab_rho}}
%\tablewidth{0pt}
\tablehead{
\colhead{Group} & \colhead{$V$ range} & \colhead{$\rho_{cnt}$\tablenotemark{a}} &
\colhead{$\rho_{fd}$\tablenotemark{a}} & \colhead{$(\rho_{cnt}-\rho_{fd})/\rho_{fd}$} & 
\colhead{N$_{star}$} & \colhead{}  }

\startdata
massive MS/post-MS & $\leq$ 10 & 1.11 & 1.34 $\times$ 10$^{-2}$ & 82.1 & 45 & \\
MS stars & 10 -- 13 & 2.86 & 5.00 $\times$ 10$^{-2}$ & 56.2 & 145 & \\
bright PMS & $\leq$ 15 & 2.23 & 3.94 $\times$ 10$^{-1}$ & 4.7 & 567 & \\
faint PMS & 15 -- 18 & 4.46 & 1.41  & 3.2 & 1817 & \\
bright non-member & above PMS locus & \nodata & 1.68 & \nodata & 1901 & \\
faint non-member & below PMS locus & \nodata & 1.74  & \nodata & 1979 & \\ \hline
\multicolumn{7}{c}{EFF profile fitting parameters} \\ \hline \hline
   & {$\mu_0$ [pc$^{-2}$]} & {$a$ [pc]} & {$\gamma$} & {$r_{\rm c}$ [pc]} & {$n_0$ [pc$^{-3}$]} & {$N_\mathrm{tot}$} \\ \hline
massive MS/post-MS &  6.80 $\pm$ 0.38    &   1.63 $\pm$ 0.04   &   5.45 $\pm$ 0.07     &   0.88 $\pm$   0.02     &    0.048$\pm$ 0.005 &   32.88  \\
MS stars & 10.18 $\pm$ 1.73    &   1.39 $\pm$ 0.18  &   2.97 $\pm$ 0.22     &   1.07 $\pm$   0.14     &    0.090 $\pm$ 0.042 &   127.15 \\
bright PMS & 10.16 $\pm$ 1.46    &  8.96 $\pm$ 1.21   &  42.34 $\pm$ 0.52     &   1.63 $\pm$   0.22     &  1.791 $\pm$ 0.094 &   126.93 \\
faint PMS  & 10.21 $\pm$ 1.63    &  7.48 $\pm$ 3.42   &  17.25 $\pm$ 1.65     &   2.16 $\pm$   0.99     &  6.494 $\pm$ 0.227 &   235.32   \\
\enddata

\tablenotetext{a}{Surface density at center and field region ($r > 10'$),
respectively. Uint: \# / $\Box ^{'}$}
\end{deluxetable}

\end{document}